\documentclass[12pt, twoside]{article}
\usepackage[semicolon,round,sort&compress]{natbib}  %
\usepackage{hyperref}
\usepackage{amsmath,amsthm, amssymb, amsfonts}
\usepackage{booktabs}
\usepackage{pdflscape}
\usepackage{calc}
\usepackage{float}
\usepackage{framed}
\usepackage{indentfirst}
\usepackage{chronology}
\usepackage[toc,page]{appendix}
\usepackage{tablefootnote}
\usepackage{subfigure}
\graphicspath{{images/}}
\usepackage{caption}
\usepackage[a4paper,width=150mm,top=25mm,bottom=25mm,bindingoffset=6mm]{geometry} 
\usepackage{filecontents}
\usepackage{algorithm}
\usepackage{algorithmic}
\usepackage{nicefrac}
\usepackage{comment}

\usepackage{graphicx}
\usepackage{verbatim}
\usepackage{multirow}
\usepackage{bbm}
\usepackage[utf8]{inputenc}
\usepackage[english]{babel}
\usepackage{mathtools}
\usepackage{natbib}
\usepackage{mathtools}

\usepackage{multicol}
\renewcommand{\bibpreamble}{\begin{multicols}{2}}
\renewcommand{\bibpostamble}{\end{multicols}}

\makeatletter
\newcommand{\distas}[1]{\mathbin{\overset{#1}{\kern\z@\sim}}}%
\newsavebox{\mybox}\newsavebox{\mysim}
\newcommand{\distras}[1]{%
  \savebox{\mybox}{\hbox{\kern3pt$\scriptstyle#1$\kern3pt}}%
  \savebox{\mysim}{\hbox{$\sim$}}%
  \mathbin{\overset{#1}{\kern\z@\resizebox{\wd\mybox}{\ht\mysim}{$\sim$}}}%
}
\makeatother

\usepackage{changepage}
\definecolor{shadecolor1}{rgb}{0.40,0.83,0.70}
\definecolor{shadecolor2}{rgb}{0.84,0.83,0.63}

 {\endMakeFramed}

\newenvironment{shaded2}{%
  \MakeFramed {\FrameRestore}}%
 {\endMakeFramed}

\title{50 shades of Bayesian testing of hypotheses\footnote{Some paragraphs of this chapter have first appeared on the author's personal blog. Section
\ref{sec:onc} mostly summarises the proposal made in \cite{kamary:mengersen:x:rousseau:2014}. The author is
very much grateful to Alastair Young for his helpful comments on earlier versions.}}
\author{{\sc Christian P.~Robert$^{1,2,3}$}\footnote{
The work of the author was partly supported in part by the French Government
under management of Agence Nationale de la Recherche as part of the ``Blanc
SIMI 1'' program, reference ANR-18-CE40-0034 and in part by the French
Government under management of Agence Nationale de la Recherche as part of the
``Investissements d’avenir'' program, reference ANR19-P3IA-0001 (PRAIRIE 3IA
Institute).  The author also gratefully acknowledges the support of l'Institut
Universitaire de France through two consecutive senior chairs.}\\
{\em $^1$CEREMADE, Université Paris Dauphine PSL},\\ 
{\em $^2$Department of Statistics, University of Warwick,}\\ 
{\em and $^3$CREST--ENSAE, Université Paris--Saclay}}
\date{}

\begin{document}
\maketitle

\begin{abstract}
Hypothesis testing and model choice are quintessential questions for statistical inference and 
while the Bayesian paradigm seems ideally suited for answering these questions, it faces difficulties
of its own ranging from prior modelling to calibration, to numerical implementation. This chapter
reviews these difficulties, from a subjective and personal perspective.
\end{abstract}

\noindent
{\bf Keywords:} 
Bayesian model selection,
BIC, DIC,
Hypothesis testing,
Improper priors,
Information criterion,
Mixtures, 
Monte Carlo, 
Posterior predictive,
Prior specification,
WAIC.

\section{Introduction}
\label{sec:intro}

The concept of hypothesis testing is somewhat inseparable from statistics as
principled hypothesis testing is unfeasible outside a statistical framework,
while testing may be the most ubiquitous and long-standing manifestation of
statistical practice, if not in theoretical statistics. The implementation of
this goal is also subject to many interpretations and controversies, as
illustrated by the recent American Statistical Association statement on the
dangers of over-interpreting $p$-values \citep{wasserstein:lazar:2016}
and other calls \citep{johnson:2013b,gelman:robert:2014,benjamin:etal:2018,blakeley:etal:2019}. In particular, testing
is a dramatically differentiating feature separating classical from Bayesian paradigms,
both conceptually and practically \cite{berger:sellke:1987,casella:berger:1987}. This
opposition will not be covered by the present chapter.

Even within the Bayesian community, testing hypotheses remains an area that is
wide open to controversy and divergent opinions, from banning any form of testing
to constructing pseudo-$p$-values. While the notion of the posterior probability
of an hypothesis appears as a ``natural" answer in a Bayesian context, there exist many issues
with that choice, from the impact of the prior modelling to the impossibility of using improper
priors, as shown by the Jeffreys-Lindley paradox \cite{lindley:1957,robert:2014}.
Furthermore, the most common binary (i.e., accept vs.~reject) outcome of an hypothesis test
appears more suited for immediate decision (if any) than for model evaluation, clashing with
the all-encompassing nature of Bayesian inference.

The literature on Bayesian hypothesis testing is huge and we can only point out to a few significant entries like
\cite{berger:1985,gelman:etal:2013,vehtari:ojanen:2012,gelman:etal:2014}. The in-depth analysis of Harold Jeffreys' 
input by \cite{ly:verhagen:wagenmakers:2016} is quite noteworthy.

\section{Bayesian hypothesis testing}
\label{sec:hippoBay}

\begin{quote}{\em
``In induction there is no harm in being occasionally wrong; it is inevitable
that we shall be. But there is harm in stating results in such a form that they do
not represent the evidence available at the time when they are stated."
} --- Harold Jeffreys (1939)\end{quote}

As a preliminary, let me point out that Bayesian hypothesis testing (or {\em model selection}, as I will use both
terms interchangeably) can be seen as a comparison of $k>1$ potential statistical models towards the selection of 
the model that fits the data ``best". A mostly accepted perspective is indeed that it does not primarily seek to
identify which model is ``true", but compares fits through marginal likelihoods or other quantities. 

A marginal likelihood (or {\em evidence}) is defined as the average likelihood function
$$m(x)=\int_\Theta L(\theta|x) \pi(\theta)\,\text d\theta$$
where $L(\theta|x)$ denotes the likelihood function attached to the sample $x$,
$\pi(\cdot)$ is the prior density and $\Theta$ the parameter
space.\footnote{This marginal sampling distribution is also called the {\em prior predictive distribution} as
the integrated standard sampling distribution with respect to the prior distribution. It can be simulated by first
generating a parameter value from the prior and second generating from the
sampling distribution indexed by this realisation of the parameter.} 
This quantity naturally includes a penalisation addressing model complexity and over-fitting, through
the averaging over the whole set $\Theta$, that is mimicked by Bayes Information (BIC) \citep{schwartz,schwarz:1978}
and Deviance Information (DIC) criteria \citep{spiegelhalter:best:carlin:linde:2002,plummer01072008}.
A fundamental difficulty with the marginal likelihood is that it exhibits a long-lasting impact of prior modeling, in that the likelihood input
does not quickly counter-balance the tail behaviour of the prior.

Each model (or hypothesis) $\mathcal M_i$ under consideration
comes with an attached triplet $(L_i(\cdot|x),\pi_i(\cdot),\Theta_i)$ $(i=1,\ldots,k)$. In addition, prior weights $\omega_i$
are characterising the prior probabilities of the models, leading to an encompassing prior
$$\pi(\theta)=\omega_1\pi_1(\theta)\mathbb I_{\Theta_1}(\theta)+\cdots+\omega_k\pi_k(\theta)\mathbb I_{\Theta_k}(\theta)$$
and the corresponding posterior probabilities
$$\pi(\mathcal M_i|x) = \dfrac{\omega_i\int_{\Theta_i} L_i(\theta_i|x)
\pi_i(\theta_i)\,\text d\theta_i}{\sum_{j=1}^k \int_{\Theta_j} \omega_j L_j(\theta_j|x)
\pi_j(\theta_j)\,\text d\theta_j}=\dfrac{m_i(x)}{\sum_{j=1}^k m_j(x)}$$
A decision-theoretic approach based on the Neyman-Pearson formalism of hypothesis testing leads to selecting the most probable model
\citep{berger:1985,lehmann:1986}. However, the strong impact of the values of the prior weights $\omega_i$ on the numerical values of
these posterior probabilities led to their removal and the construction of the Bayes factors \citep{wrinch:jeffreys:1919,haldane:1932,jeffreys:1939}, 
comparing the marginal likelihoods between models $(L_i,\pi_i)$ and $(L_j,\pi_j)$ by the ratio
\begin{equation}\label{eq:BF}
\mathfrak B_{ij} = \dfrac{\int_{\Theta_i} L_i(\theta_i|x)
\pi_i(\theta_i)\,\text d\theta_i}{\int_{\Theta_j} L_j(\theta_j|x)
\pi_j(\theta_j)\,\text d\theta_j}\end{equation}
which amounts to selecting among models the model with the highest marginal. This is also advocated as an even weighting of both models 
in \cite{jeffreys:1939}, but I see little justification for this choice, especially when considering multiple model selection
with possibly embedded models.

\begin{figure}
\centerline{\includegraphics[width=.6\textwidth]{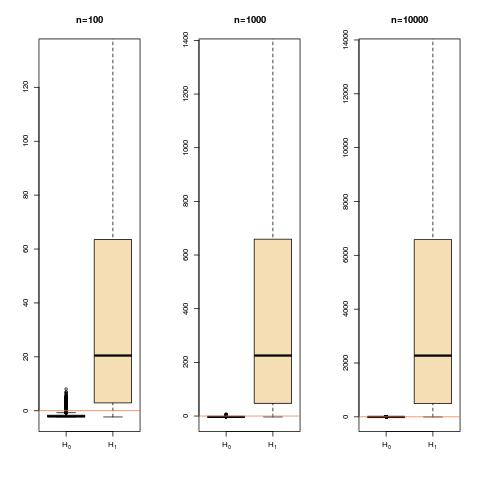}}
\caption{\label{fig:constor}\small Example of a normal ${\bar X}_n\sim\mathcal{N}(\mu,\nicefrac{1}{n})$ model 
when $\mu\sim\mathcal{N}(0,1)$ under $H_1$ and $\mu=0$ under $H_0$, leading to the Bayes factor 
$\mathfrak{B}_{10}=(1+n)^{-\nicefrac{1}{2}}\exp\{n^2{\bar x}^2_n/2(1+n)\}$. The three panels represent 
the variability of $\log \mathfrak{B}_{10}$ over 250 replicas under the null {\em (left)} and the alternative {\em (right)} hypotheses,
for three different sample sizes $n$. In the later case, the replications are generated from the prior predictive distribution.
This experiment is supporting the statement about the {\em consistency} of the Bayes factor, namely that the true model is selected by the Bayes
factor when the sample size $n$ goes to infinity, since the Bayes factor $\mathfrak{B}_{10}$ accumulates at zero under $H_0$ and diverges under $H_1$.}
\end{figure}

The coexistence of both notions---posterior probability versus Bayes factor---exhibits a tension between using (i) posterior probabilities 
as justified by binary loss functions but depending on subjective prior weights that prove difficult to
specify in most settings, especially when comparing many hypotheses, and (ii) Bayes factors 
that eliminate this dependence but escape a direct connection with a posterior distribution, unless the
prior weights are themselves integrated within the loss function. A further difficulty attached to the
Bayes factors is that they face a delicate interpretation (or calibration) of the strength of their support of
a given hypothesis or model, even when shown to be consistent in the sample size
\citep{berger:ghosh:mukhopadhyay:2003,dass:lee:2004}, as illustrated by Figure
\ref{fig:constor}. That is, under fairly generic conditions on the priors, the
Bayes factor $\mathfrak B_{ij}$ will diverge to infinity or to zero when the
sampling model is $\mathfrak M_i$ or $\mathfrak M_j$, respectively, as the sample 
size grows to infinity \citep{chib:kuffner:1996,ohagan:1997,johnson:2008,moreno:giron:casella:2010}.
The differentiation between the simulated values of the Bayes factors under the
null model and under the alternative model is getting more pronounced in this figure
as the sample size grows.  As functions of the data $x$, both notions are such
that their calibration and a necessary variability assessment seem to require
(frequentist or posterior predictive) simulations under both hypotheses, which
clashes with the Bayesian paradigm. However, posterior probabilities face a
similar difficulty if one wants to avoid interpreting them as $p$-value
substitutes or probabilities of selecting the ``true" model.  since they only
report of respective strengths of fitting the data $x$ to the models under comparison.

\begin{quote}{\em
``Bayes factors also suffers from several
theoretical and practical difficulties. First, when improper prior
distributions are used, Bayes factors contains undefined constants and takes
arbitrary values (...) Second, when a proper but vague prior distribution with
a large spread is used to represent prior ignorance, Bayes factors tends to
favour the null hypothesis. The problem may persist even when the sample size
is large (...) Third, the calculation of Bayes factors generally requires the
evaluation of marginal likelihoods. In many models, the marginal likelihoods
may be difficult to compute.”
} -- Yong Li, Tao Zen, and Jun Yu (\citeyear{li:zen:yu:2014})
\end{quote}

Among other difficulties inherent to the use of the Bayes factor, and as illustrated by the above quote,
let me mention an impossibility to ascertain simultaneous misfits to all proposed
models (unless a non-parametric alternative is added as in \citealp{holmes:etal:2015}) or to detect outliers, that is a subset of the
dataset that does not fit a particular model. Another pressing issue I will not address here is the challenging numerical computation 
of marginal likelihoods in most settings. While numerous proposals have been made \citep[see, e.g.,][]{chen:shao:ibrahim:2000,robert:casella:2004,
chopin:robert:2010,friel:pettitt:2008,robert:marin:2008,marin:robert:2010,friel:wyse:2012}, there is no universal solution and new settings 
require a careful design of the numerical apparatus producing the evidence. A last difficulty worth pointing out is the 
strong dependence of posterior probabilities and Bayes factors on conditioning
statistics, which undermines their validity for model assessment.  This issue
was exhibited when considering ABC (Approximate Bayesian computation, see
\citealp{sissonfanbeaumont:2018}) model choice as the lack of inter-model
sufficiency may drastically alter the value of a Bayes factor and even its
consistency \citep{didelot:everitt:johansen:lawson:2011,
robert:cornuet:marin:pillai:2011,marin:pillai:robert:rousseau:2011}.

\section{Improper priors united against hypothesis testing}
\label{sec:improper}

Hypothesis testing sees a glaring discontinuity occur in the valid use of improper (infinite mass) priors since calling on these is not
directly justified in most testing situations, leading to many alternative and {\em ad hoc} solutions, where data is either used twice or split
in artificial ways. This is most unfortunate in that the remainder of Bayesian analysis accommodates rather smoothly the extension from proper (that is,
true probability) prior distributions to improper (that is, $\sigma$-finite) prior distributions, which is in particular most helfpul in closing the
range of Bayesian procedures. The fundamental reason for the difficulty in the testing context is the necessity to define a prior distribution for each
model under comparison, {\em independently} of the other priors. This means there is no rigorous way of defining a model-by-model normalisation of 
these priors when they are $\sigma$-finite. (Principled constructions of reference priors for testing have been investigated, see for instance
\citealp{bayarri:garcia:2007} and \citealp{bayarri:garcia:2008}. Their proposal is based on symmetrised versions of the Kullback-Leibler divergence 
$\kappa$ between the null and alternative distributions, transformed into a prior that looks like an inverse power of $1+\kappa$,
with a power large enough to make the prior proper.)

A first difficulty with the proposed resolutions of the improper conundrum stands
with the choice (already found in \citealp{jeffreys:1939}) of opting for
{\em identical priors} on the parameters present in both models \citep{berger:pericchi:varshavsky:1998}, which amounts in
endowing them with the same meaning. The following argument of then using {\em the same prior}, whether or not proper, then eliminates the need for
a normalising constant. (Note that the Savage--Dickey approximation of the marginal likelihood relies on this assumption and operates only under 
the alternative hypothesis, see \citealp{verdinelli:wasserman:1995,marin:robert:2010b}.)

A second difficulty is attached to the pseudo-Bayes factors proposed in the 1990's \citep{ohagan:1995,berger:pericchi:varshavsky:1998}, where a fraction of the
data is used to turn all improper priors into proper posteriors and these posteriors are used as new ``priors" on the remaining fraction. On the one
hand this is a nice bypass of the normalisation constant issue and it enjoys consistency properties as the sample size of the remaining fraction grows
to infinity. Furthermore, it does not use the data {\em twice} as \citep{aitkin:1991,aitkin:2010}. On the other hand, being a leave-one-out approach, it does not
qualify as a Bayesian procedure proper and the different manners to average over all possible choices of the normalising sample lead to different numerical
answers, while there exist cases when this division proves impossible.

Alternative approaches have advocated the use of score functions \citep{hyvarinen:2005,gutmann:hyvarinen:2012,li:zen:yu:2014,dawid:musio:2015,shao:etal:2017}
to overcome the issue with improper priors. While quite sympathetic to this perspective, I will not cover this aspect in this chapter.

While the deviance information criterion (DIC) of \citep{spiegelhalter:best:carlin:linde:2002,spiegelhalter14} 
remains quite popular for model comparison, its uses of the posterior expectation of the log-likelihood function, 
meaning that the data is used twice, as in \cite{aitkin:2010}, and of a plug-in term, make
it disputable, as discussed in \cite{spiegel:disc:14} and prone to conflicting
interpretations, as shown in \cite{celeux:forbes:robert:titterington:2006} for
mixtures of distributions. A related approach with stronger theoretical backup
is the widely applicable information criterion (WAIC) of \cite{watanabe:2010,watanabe:2018},
which is asymptotically equivalent to the average Bayes generalization and cross validation losses. The
fundamental setting of WAIC is one where both the sampling and the prior
distributions are different from respective ``true” distributions. This requires using a
tool towards the assessment of the discrepancy when utilising a specific pair of such
distributions, especially when the posterior distribution cannot be
approximated by a Normal distribution.  The WAIC is supported for the determination of
the ``true” model, in opposition to AIC and DIC. In addition, it escapes the ``plug-in" sin and is handling mixture models.

\section{The Jeffreys--Lindley paradox}
\label{sec:prdX}

\begin{quote}{\em
``The weight of Lindley's paradoxical result (...)
burdens proponents of the Bayesian practice".
} -- Frank Lad (\citeyear{lad:2003})
\end{quote}

The Lindley paradox (or Jeffreys--Lindley paradox) is due to
\cite{lindley:1957} pointing out an irreconciliable divergence between
the classic and Bayesian procedures when testing a point null hypothesis.
(This property was briefly mentioned in
\citealp{jeffreys:1939}, V, \S5.2, although in their scholarly review of the paradox, \citealp{wagenmakers:ly:2021} stress 
that it plays a central role in his construction of Bayesian hypothesis testing.) The paradox is that, regardless of the prior 
choice, the Bayes factor against the null hypothesis converges to zero with the sample size when the associated $p$-value 
remains constant. For instance, when testing the nullity of a Normal mean, 
$$
\bar x_n \sim \mathcal{N}(\theta,\sigma^2/n)\,,\quad
H_0:\,\theta=\theta_0\,,
$$
with the following $H_1$ prior, $\theta\sim\mathcal{N}(\theta_0, \sigma^2)$, the Bayes factor is
$$
\mathfrak{B}_{01}(t_n) = (1+n)^{1/2}\,\exp\left(-n t_n^2/2[1+n]\right)\,, 
$$
where $t_n=\sqrt{n}|\bar x_n - \theta_0|/\sigma$. When setting $t_n$ to a fixed value, it
converges to infinity with $n$.

\begin{quote}{\em
`` The Jeffreys--Lindley paradox exposes a rift between Bayesian and
frequentist hypothesis testing that strikes at the heart of statistical inference."
} -- Eric-Jan Wagenmakers and Alexander Ly (\citeyear{wagenmakers:ly:2021})
\end{quote}

Since the apparent paradox of ``always" accepting the null hypothesis can be reformulated in terms of a prior variance going to infinity, 
it also relates with the difficulty in using Bayes factors and improper priors, in that the posterior mass of the region with non-negligible
likelihood goes to zero as the variance increases \citep{robert:2013}. This is thus a coherent framework in that the only remaining item of
information is that the null hypothesis could be true! (Note also that the paradox can be circumvoluted by replacing the point null hypothesis
with an interval substitute in order for a single proper or improper prior to be used, see also \citealp{robert:1993b}.)

The opposition between frequentist and Bayesian procedures is not a surprise either. The former relies solely on the point-null hypothesis $H_0$ 
and the corresponding sampling distribution, while the latter opposes $H_0$ to a (predictive) marginal version of $H_1$. Furthermore, the fact that
the statistic $t_n$ remains constant (or equivalently that the Type I error remains constant) is incorrect. The rejection bound
cannot be a constant multiple of the standard error as the sample size $n$ increases, as demonstrated by \cite{jeffreys:1939} and discussed in details by
\cite{wagenmakers:ly:2021}. Let me conclude by mentioning the case for specific priors isolating the null from the alternative hypotheses, as in
\cite{consonni:veronese:1987,johnson:rossell:2010} and \cite{consonni:etal:2013}.

\section{Posterior predictive $p$-values}
\label{sec:poprep}

Once a Bayes factor $\mathfrak B_{01}$  is computed, one need assess its
strength in supporting one of the hypotheses, if any. In my opinion, the much vaunted Jeffreys’ (1939) scale has 
very little validation as it is absolute (i.e., with no dependence on the model, the sample size,
the prior). Following earlier proposals in the literature
\citep{box:tiao:1992,garciadonato:chen:2005,geweke:amisano:2008}, an evaluation
of this strength within the issue at stake, i.e., the comparison of two hypotheses (or models), can be
based on the predictive distributions. That is, the {\em likelihood} of observing $\mathfrak B_{01}(x^\text{obs})$, the observed dataset,
is evaluated under these distributions
$$\mathbb P_0^\text{pred}(\mathfrak B_{01}(X^\text{pred})\ge \mathfrak B_{01}(x^\text{obs})|x^\text{obs})$$
and
$$\mathbb P_1^\text{pred}(\mathfrak B_{01}(X^\text{pred})\le \mathfrak B_{01}(x^\text{obs})|x^\text{obs})$$
the probabilities being computed under models $\mathfrak M_0$ and $\mathfrak M_1$, respectively.\footnote{Using a single encompassing
predictive is possible, but this distribution depends on the usually arbitrary prior probabitlies of both models.}

While most authors (like García-Donato
and Chen, 2005) consider this should be the {\em prior} predictive distribution, I agree with \cite{gelman:etal:2013}
that using the push-forward image by $\mathfrak B_{01}(\cdot)$ of the posterior predictive distribution 
\begin{equation}\label{eq:pospre}
\pi(x^\text{rep}|x^\text{obs}) = \int p(x^\text{rep}|\theta) \pi(\theta|x^\text{obs})\,\text d\theta
\end{equation}
(where $x^\text{obs}$ denotes the observed dataset and $x^\text{rep}$ an artificial replication or a running variate)
is more relevant. Indeed, by exploiting the information contained in the data (through the posterior),
\eqref{eq:pospre} concentrates on a region of relevance in the parameter
space(s), which is especially interesting in weakly informative settings,
despite ``using the likelihood twice".\footnote{
This double use can possibly be argued for or against, once a data-dependent loss function is built, but the potential
for over-fitting must be investigated on its own, globally or model by model. However, the above probabilities can also be
perceived as producing an estimator of the posterior loss.}
Furthermore, \eqref{eq:pospre} evaluate the behaviour of the Bayes factor for values of $x$ that are similar to the original observation, 
provided the posterior predictive fits the data well enough. Note also that, under this approach, issues of indeterminacy linked with improper priors 
are not evacuated, since the Bayes factor remains indeterminate, even with a well-defined predictive.
Note further that, even though probabilities of errors of type I and errors of type
II can be computed, they fail to account for the posterior probabilities of
both models. (This is a potentially delicate issue with the solution of García-Donato and
Chen, 2005.) A nice feature is that the predictive distribution of the Bayes
factor can be computed even in complex settings when ABC \citep{sissonfanbeaumont:2018} need be used.

\begin{quote}{\em
``If the model fits, then replicated data generated under the model should look similar to the observed data.
The observed data should look plausible under the posterior predictive distribution. This is really
a self-consistency check: an observed discrepancy can be due to model misfit or chance."
} -- Andrew Gelman {\em et al.} (2013) \end{quote}

In {\em Bayesian Data Analysis} (2013, Chapter 6), based on the choice of a measure of discrepancy $T(\cdot,\cdot)$, the
authors (strongly) suggest replacing the classical $p$-value
$$
p(x^\text{obs}|\theta) = \mathbb P(T(X^\text{rep},\theta)\ge T(x^\text{obs},\theta)|\theta)
$$
with a Bayesian alternative (or Bayesian posterior $p$-value)
$$
\mathbb P(T(X^\text{rep},\theta)\ge T(x^\text{obs},\theta)|x^\text{obs})\,.
$$
Extremes $p$-values indicate a poor fit of the model, with the usual caution applying about setting golden bounds such as $0.05$
or $0.99$. There are however issues with the implementation of this approach, from deciding on which aspect of the data or of the model is 
to be examined, i.e., the choice of the discrepancy measure $T$, to its calibration.

\section{A modest proposal}
\label{sec:onc}

Given this rather pessimistic perspective on Bayesian testing, one may wonder
at the overall message of this chapter besides the not-particularly useful ``it
is complicated". Given the difficulty in moderating the impact of the prior modelling
in ways more useful than a sheer assurance of consistency, my preference leans towards
a proposal that feels more estimation-based than testing-based, and that is leaning more 
towards quantification than towards binary decision.
In \cite{kamary:mengersen:x:rousseau:2014}, we have sketched the basis of a novel approach
that advocates the replacement of the posterior probability of a model or of an hypothesis 
with the posterior distribution of the weights of a mixture of the models under comparison. 
That is, given two Bayesian models under comparison,
$$
\mathfrak{M}_1:x\sim f_1(x|\theta_1)\,,\theta_1\sim\pi_1(\theta_1) \text{
versus } \mathfrak{M}_2:x\sim f_2(x|\theta_2)\,,\theta_2\sim\pi_2(\theta_2)
$$
we propose to estimate the (artificial) mixture model
\begin{equation}\label{eq:K2R2}
\mathfrak{M}_{\alpha}:x\sim\alpha f_1(x|\theta_1) + (1-\alpha) f_2(x|\theta_2)
\quad 0\le \alpha\le 1
\end{equation}
and in particular to derive the posterior distribution of $\alpha$. This (marginal) posterior can then
be exploited to assess the better fit and if need be to achieve a decision, either by assessing tail probabilities
that $\alpha$ is close to $0$ or $1$, or by calibrating a bound from the prior or posterior predictives. 
In most settings, this approach can indeed be easily calibrated
by a parametric bootstrap experiment providing a posterior distribution of $\alpha$
under each of the models under comparison. The prior predictive error can
therefore be directly estimated and drive the choice of a decision cut-off on the
tails of $\alpha$, if need be.

\begin{figure}[!h]
\includegraphics[width=.33\textwidth]{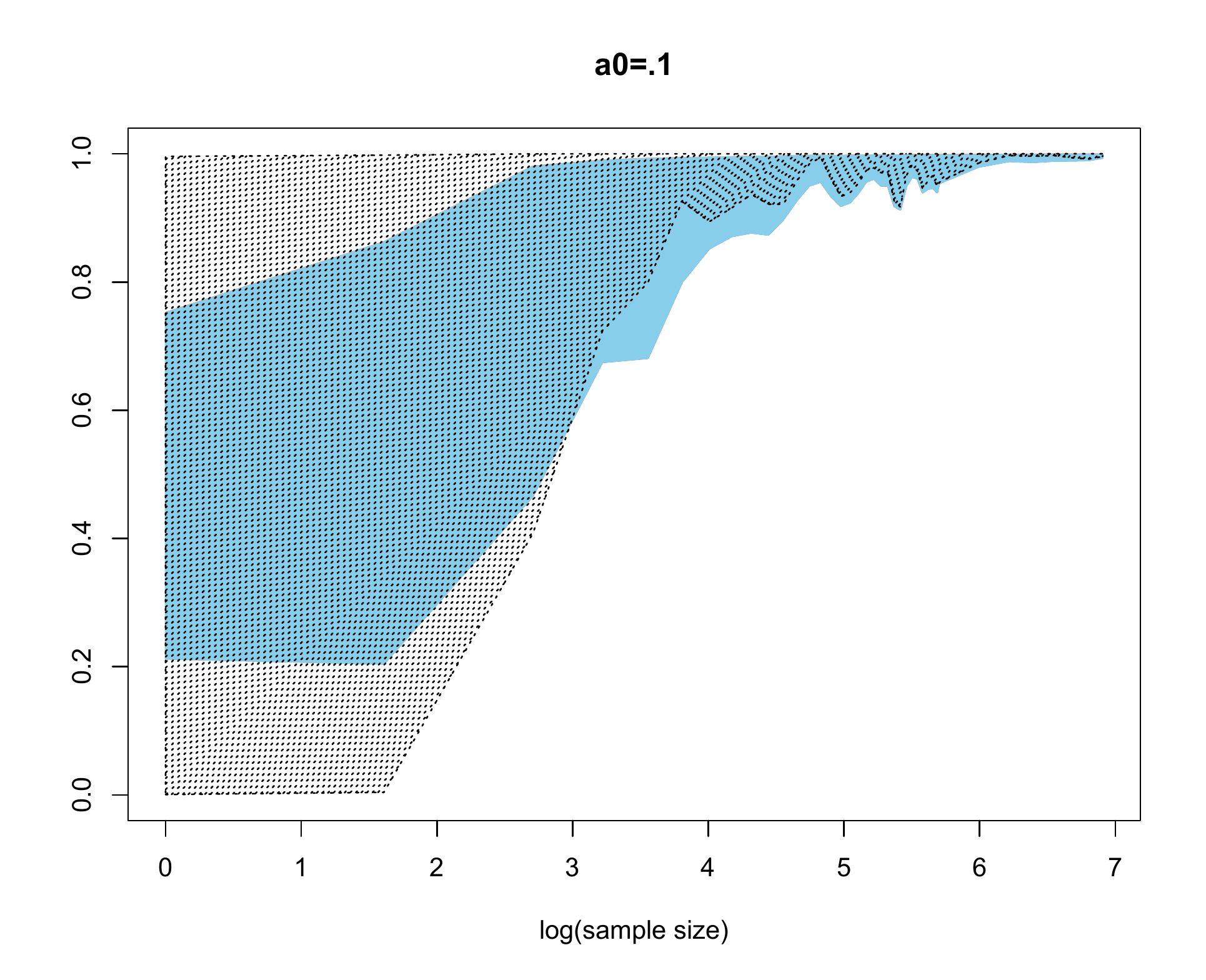}\includegraphics[width=.33\textwidth]{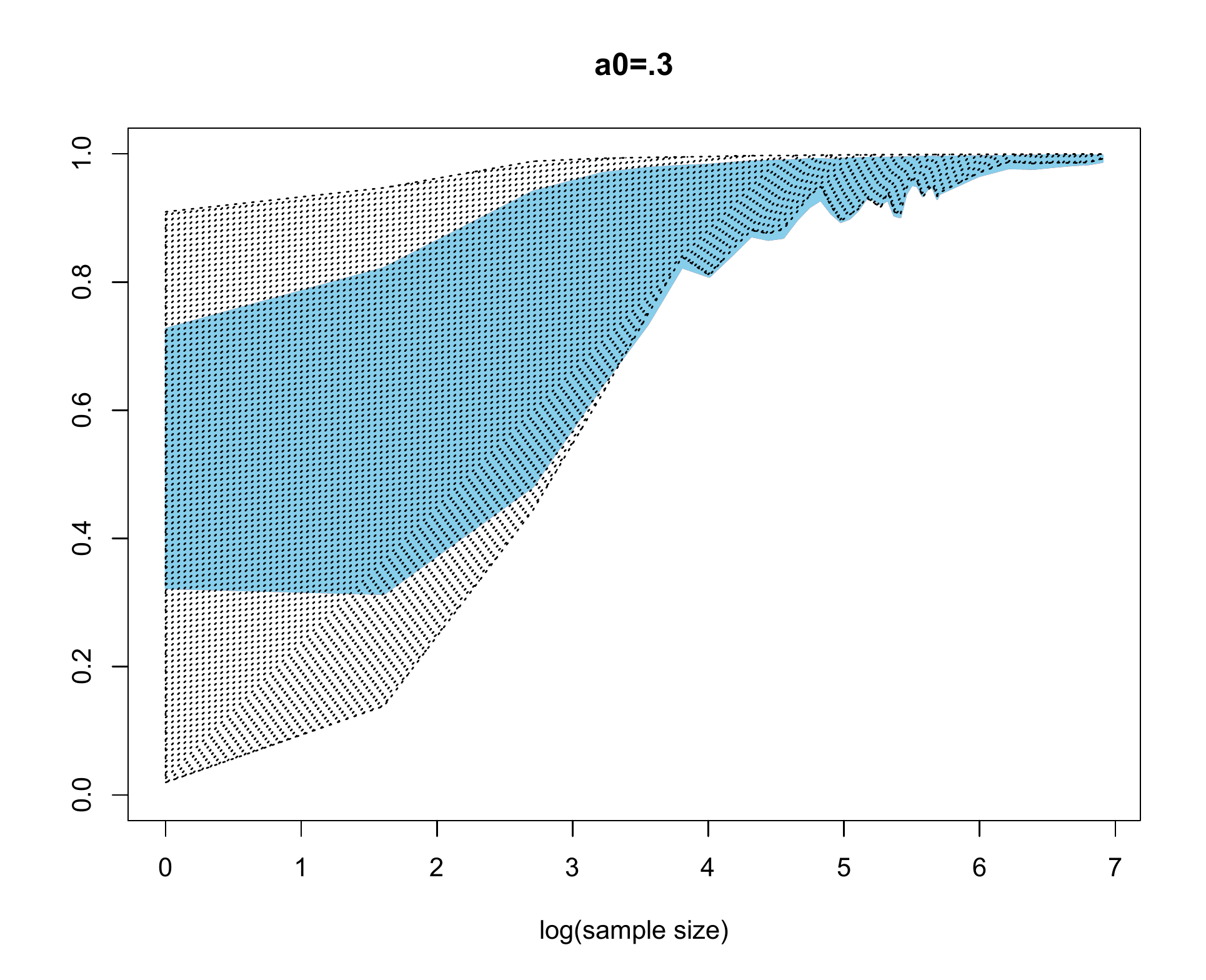}\includegraphics[width=.33\textwidth]{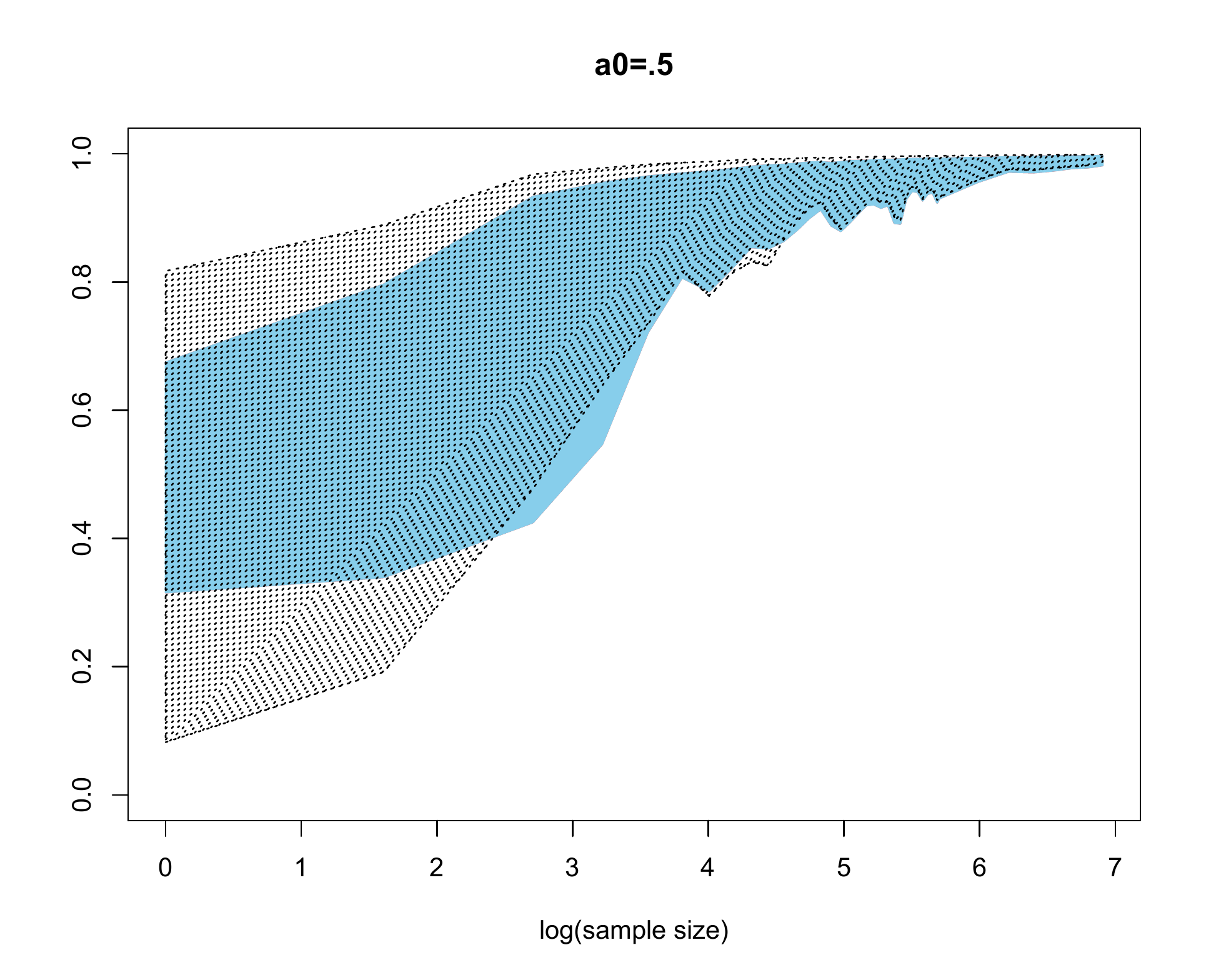}
\caption{\small Posterior means {\em (sky-blue)} and medians {\em (grey-dotted)} of the posterior
distributions on $\alpha$, displayed over $100$ simulations of Poisson $\mathcal{P}(4)$ datasets for sample sizes $n$ between 1 and 1000. The
shaded and dotted areas indicate the range of the estimates. Each plot corresponds to a $\mathcal{B}e(a_0,a_0)$ prior on $\alpha$ 
and each posterior approximation is based on $10^4$ MCMC iterations. This experiment is supporting the statement about the concentration of
the posterior on the mixture weight near the upper limit $1$, in agrement with the correct model. {\em (Source: \citealp{kamary:mengersen:x:rousseau:2014})}}
\label{pgb2}
\end{figure}

Consider for instance a simple example in \cite{kamary:mengersen:x:rousseau:2014} where $f_1(\cdot|\theta_1)$
corresponds to a Poisson $\mathcal{P}(\theta_1)$ distribution and $f_2(\cdot|\theta_1)$ to a Geometric $\mathcal{G}eo(\theta_2)$
failure distribution, both parameterised in terms of their mean. A mixture
$$ \alpha \mathcal{P}(\lambda) +(1-\alpha) \mathcal{G}eo(\nicefrac{1}{1+\lambda})$$
can then be proposed, with the same parameter $\lambda>0$ behind both components. Further, 
\cite{kamary:mengersen:x:rousseau:2014} show that the non-informative prior $\pi(\lambda)=\nicefrac{1}{\lambda}$ can be used
in this setting, whatever the sample size $n$. Figure \ref{pgb2} demonstrates the concentration of
the posterior on $\alpha$ around $1$ as $n$ increases when the data generating distribution is a $\mathcal P(\lambda)$ distribution.

\begin{figure}[!h]
\centerline{\includegraphics[width=.4\textwidth]{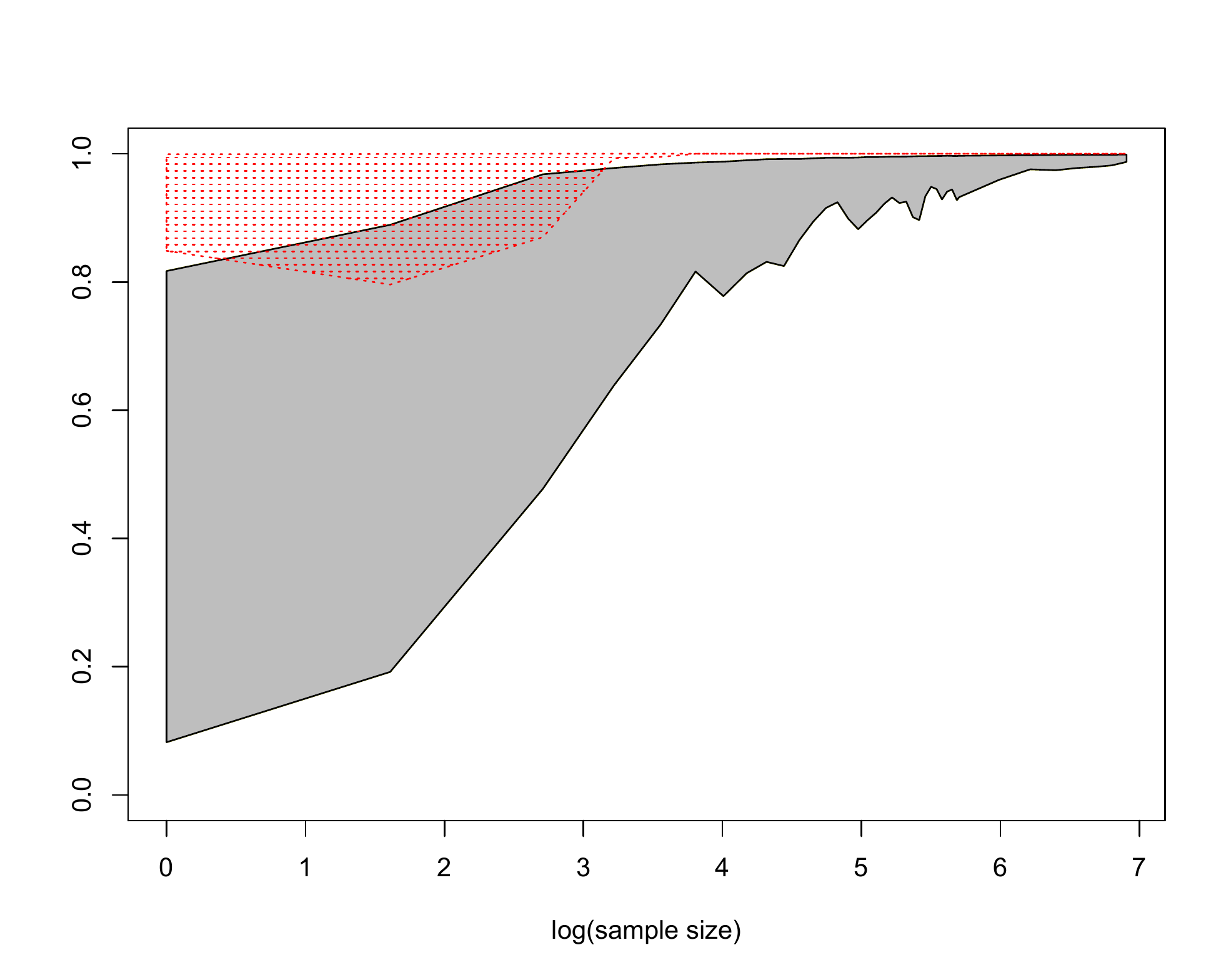}}
\caption{\label{fig:mousee}\small Comparison between the ranges of $\mathbb{P}(\mathfrak{M}_1|x)$ {\em (red dotted
area)} and of the posterior medians of $\alpha$ for $100$ simulations of Poisson $\mathcal{P}(4)$ datasets with sample sizes $n$
ranging from $1$ to $10^3$. {\em (Source: \citealp{kamary:mengersen:x:rousseau:2014})}}
\end{figure}

In this example, the exact (standard) Bayes factor comparing the Poisson to the Geometric models is given by
$$
\mathfrak{B}_{12} =n^{n \bar{x}_n}\prod_{i=1}^n
x_i!\,  {\Gamma \left(n+2+n \bar{x}_n\right)}\big/{\Gamma(n+2)}.
$$
when using the {\em same} 
improper prior on the parameter $\lambda$  \citep{degroot:1973,degroot:1982,berger:pericchi:varshavsky:1998}.
The posterior probability of the Poisson model is then derived as
$$
\mathbb{P}(\mathfrak{M}_1|x)=\frac{\mathfrak{B}_{12}}{1+\mathfrak{B}_{12}}
$$
when adopting (without much of a justification) identical prior weights on both models. Figure \ref{fig:mousee} compares the concentration of
the posterior probabiliy and of the posterior median on $\alpha$ around $1$ as $n$ increases.

One may object that the mixture model \eqref{eq:K2R2}  is neither one nor the other of the models under comparison,
but it includes both of these at the boundaries, i.e., when $\alpha=0,1$. Thus, if we use a 
prior distribution on $\alpha$ that favours neighbourhoods of $0$ and $1$, albeit avoiding atoms at these values,
we should be able to witness the posterior concentrate near $0$ or $1$, depending on which model is true (or fits
the data better).  This is indeed the case: as shown in \cite{kamary:mengersen:x:rousseau:2014}, for any given Beta 
prior on $\alpha$, we observe a higher and higher concentration at the right boundary as the sample size $n$ increases,
as illustrated by Figure \ref{pgb2}.

In our opinion, this (novel) mixture approach offers numerous advantages.
First, achieving a decision while relying on a Bayesian estimator of the weight
$\alpha$ or on its posterior distribution, rather than on
the posterior probability of the associated model lifts the embarassing
need of specifying almost invariably artificial prior probabilities on the models,
as discussed above. This is also a most Bayesian choice, replacing an unknown
quantity with a probability distribution. It is however not addressed by a
classical Bayesian approach, even though those probabilities linearly impact
the posterior probabilities and their indeterminacy is often brought
forward to promote the alternative of using instead the Bayes factor. In the mixture estimation setting, prior
modelling only involves selecting a prior on $\alpha$, for instance a Beta $\mathcal
B(a,a)$ distribution, with a wide range of acceptable values for the hyperparameter $a$.
While its value obviously impacts the posterior distribution of $\alpha$, it can be argued
that it still induces an accumulation of the posterior mass near $1$ or $0$, i.e.,
favours the most favourable or the true model over the other one, and a
subsequent sensitivity analysis on the impact of $a$ is elementary to implement

The interpretation of the estimator of $\alpha$ is 
furthermore at least as ``natural" as handling the corresponding posterior probability, while 
avoiding the rudimentary zero--one loss function. The quantity $\alpha$ and its posterior
distribution provide a Bayesian measure of proximity to either model for the data at
hand, which is a most reasonable measure of fit, while being also interpretable as a
propensity of the data to stand with (or to stem from) these models.
This representation further allows for alternative perspectives on testing and
model choices, through the notions of predictive tools,  cross-validation
\citep{vehtari:lampinen:2002,vehtari:ojanen:2012,gelman:etal:2013}, and
information indices like WAIC \citep{watanabe:2010,watanabe:2018}. They may further relate to sparsity
priors like the horseshoe priors \citep{carvalho:polson:scott:2009} in Bayesian variable selection, which
is examined in \cite{kamary:mengersen:x:rousseau:2014}.

From a computational perspective, the highly challenging computation of the marginal likelihoods is
absent fom this approach, since standard algorithms are available for Bayesian mixture estimation
\citep{celeux:etak:2020}. In addition, the extension of this perspective to a finite collection of models 
under comparison is immediate, as this modelling simply expands the mixtureinto a larger number of components.
This approach further allows to involve all models at once rather than engaging in
many pairwise comparisons, thus eliminating the least likely models by
simulation. This is much more efficient than in the alternative reversible
jump strategies \citep{green:1995}. Note as a side remark that the (conceptually and computationally) challenging
issue of ``label switching” \citep{stephens:2000b,jasra:holmes:stephens:2005} 
attached with most mixture models does not appear in this particular context, since
mixture components are then not exchangeable. In particular, the mixture representation involves
neither a Bayes factor nor a posterior probability and hence it bypasses the difficulty of 
exploring all modes of the
posterior distribution. Thus, this perspective is solely focussed on estimating the
parameters of a mixture model where all components are identifiable.

From an inferential perspective, we deem that the posterior distribution of $\alpha$ evaluates more thoroughly the strength
of the support for a given model than the single-figure outcome of a Bayes factor derivation. The valuable variability of the posterior
distribution on $\alpha$ allows for a more thorough assessment of the strength of the data-support of one model against the other.
In a related manner, an additional and crucial feature missing from more traditional Bayesian answers is that a mixture 
model also acknowledges the significant possibility that, for a finite dataset, both models or none could be acceptable.
Also significantly, while standard (proper and informative) prior modelling can be
painlessly reproduced in this mixture setting, we stress that some (if not all) improper priors
can be managed via this approach, provided both models under comparison are first
reparametrised towards common-meaning and cross-model common parameters, as for instance
with location or/and scale parameters. In the special case when {\em all} parameters can
be made common to both models\footnote{While this may sound like an extremely
restrictive requirement in a traditional mixture model, let us stress here that
the presence of common parameters becomes quite natural within a testing
setting. To wit, when comparing two different models for the same data, moments
are defined in terms of the observed data and hence should enjoy the same meaning across
models. Reparametrising the models in terms of those common-meaning moments
does lead to a mixture model with some and maybe all common parameters. We thus
advise the use of a common parametrisation, whenever possible.} the mixture
model reads as
$$
    \mathfrak{M}_\alpha\,:\, x\sim \alpha f_1(x|\theta) + (1-\alpha)f_2(x|\theta) \,.
$$
For instance, if $\theta$ is a location parameter, a flat prior can be used
with no foundational difficulty, in opposition to the testing case. Following this
line of argument, we feel that using the {\em same} parameters or
some identical parameters on both components is an essential feature of this
reformulation of Bayesian testing, as it highlights the fact that the
opposition between both components of the mixture is not an issue of
enjoying different parameters, but quite the opposite: as further stressed
below, this or even those common parameter(s) is (are) nuisance parameters that
need be integrated out (as they also are in the traditional Bayesian approach
through the computation of the marginal likelihoods).

\section{Conclusion}

This coverage of the different options for conducting Bayesian testing and Bayesian model choices is
obviously subjective and other authors \cite{gelman:2018,held:2018,magnusson:2020,wagenmakers:2018,vandeshoot:2021}
in the field differ in their assessment of how this aspect of Bayesian inference should be conducted (or prohibited).
However, my exposition hopefully reflects on the complexity of the task at hand and on the necessity to avoid ready-made
solutions as those based on binary losses. Whatever the perspective adopted, it should always be calibrated through
simulated synthetic data from the different models under comparison.

\small
\hyphenation{Post-Script Sprin-ger}


\end{document}